\documentclass[twocolumn,showkeys, superscriptaddress]{revtex4-2}

\usepackage{mathpazo}
\usepackage[T1]{fontenc}
\usepackage[utf8]{inputenc}
\usepackage{xcolor}
\usepackage{mathtools}
\usepackage{amsmath}
\usepackage{amssymb}
\usepackage{graphicx}
\usepackage[bookmarks=false,
 breaklinks=true,pdfborder={0 0 1},backref=false,colorlinks=true]
 {hyperref}
\hypersetup{
 unicode=false,pdftoolbar=true,pdfmenubar=true,pdffitwindow=false,pdfstartview={FitH},linkcolor=red,citecolor=blue}

\begin{document}

\title{Dynamics of localized solutions in three core coupled waveguides with
quasi-periodic nonlinearity}

\author{Bruno M. Miranda}
\email{bruno.martins@ifg.edu.br}
\affiliation{Instituto de Física, Universidade Federal de Goiás, Goiânia, Goiás,
74.690-900, Brazil}

\author{Ardiley T. Avelar }
\affiliation{Instituto de Física, Universidade Federal de Goiás, Goiânia, Goiás,
74.690-900, Brazil}

\author{Wesley B. Cardoso }
\affiliation{Instituto de Física, Universidade Federal de Goiás, Goiânia, Goiás,
74.690-900, Brazil}

\author{Dionisio Bazeia}
\affiliation{Departamento de Física, Universidade Federal da Paraíba, João Pessoa,
Paraíba, Brazil}

\begin{abstract}
In this paper we investigate the behavior of localized solutions,
specifically solitons, in a system of three coupled waveguides. The
nonlinearity is modeled by a quasi-periodic modulation influencing
the interaction between the waveguides. We analyze the evolution of
the soliton profiles and their dynamics under varying modulation parameters,
highlighting distinct behaviors such as attraction and repulsion among
solitons. Our findings reveal that the system exhibits complex behaviors,
depending on the interplay between the quasi-periodic modulation and
the waveguide parameters. The study contributes to understanding the
impact of quasi-periodic nonlinearity on soliton dynamics in coupled
waveguide systems, laying the groundwork for potential applications
in nonlinear optics and photonic devices.
\end{abstract}


\maketitle

\section{Introduction} \label{intro}

The intriguing physics of nonlinear systems, such as nonlinear optics
\citep{Kivshar03}, Bose-Einstein condensates (BECs) \citep{Pitaevski03},
plasmas \citep{Chen_PRL76}, hydrodynamics \citep{Osborne_PRL91},
photorefractive phenomena \citep{Trillo01}, nematic liquid crystals
\citep{Assanto03}, and electrical transmission lines \citep{Remoissenet13},
underscores the importance of localized phenomena like bright and
dark solitons, breathers, and rogue waves. Optical solitons, in particular,
exhibit the unique capability to propagate over extended distances
without attenuation while maintain their shape, making them strong
candidates for primary carriers in telecommunications \citep{Haus_RMP96}.
Consequently, a significant amount of theoretical and experimental
research has focused on understanding the dynamics of solitons in
optical waveguides, particularly in silica fibers, across various
scenarios. These studies are primarily motivated by the potential
implementation of soliton-based optical communication systems.

The interaction of solitons, which can manifest as elastic collisions
or complex bound states, is a fundamental aspect of nonlinear dynamics
with significant implications across various scientific and technological
domains \citep{Adamatzky12}. In nonlinear optics, soliton interactions
enable the design of advanced signal processing techniques and the
development of all-optical logic gates, which are essential for high-speed
telecommunications \citep{Kanna_PRE03}. Within Bose-Einstein condensates
(BECs), soliton collisions provide insights into quantum coherence
and matter-wave interference phenomena \citep{Nath_PRA07,Becker_NP08},
offering potential applications in quantum information processing
and precision measurement. Furthermore, the study of soliton interactions
in plasmas and hydrodynamics enhances our understanding of wave propagation
in nonlinear media, which is crucial for applications in controlled
nuclear fusion and oceanography \citep{Zabusky_PRL65,Ablowitz_PRE12}.
The ability to precisely manipulate soliton interactions thus opens
new avenues for innovation in fields ranging from photonics to quantum
technologies.

Indeed, soliton collisions have garnered significant attention from
researchers due to the fascinating phenomena associated with scattering
mechanisms \citep{Yajima_PTP76,Goorjian_OL92,Mollenauer_OL95,Fukushima_PLA95,Yang_PRL00,Tan_PRE01,Holmer_CMP07,Dmitriev_PRE01,Dmitriev_PB02,Dmitriev_PRE02,Craig_PF06,Zhu_PRE07,Zhu_PRL08,Goodman_C08,Dmitriev_PRE08,Cardoso_PLA10,Cardoso_PRE12,Teixeira_PLA16,Cardoso_ND19,Vergara_CSF21,Cardoso_ND21,Ramakrishnan_ND24,Xu_ND24,Han_OLT23}.
Specifically, the formation and interaction of so\-nic-Lang\-muir
solitons were studied in Ref. \citep{Yajima_PTP76}. In Ref. \citep{Goorjian_OL92},
femtosecond electromagnetic soliton propagation and collision were
observed. Additionally, it was experimentally discovered that soliton-soliton
collisions in wavelength division multiplexing significantly altered
the polarization states of the solitons in Ref. \citep{Mollenauer_OL95}.
The soliton scattering by impurity potentials in a perturbed sine-Gordon
equation was numerically investigated in Ref. \citep{Fukushima_PLA95}.
The collision of orthogonally polarized, equal-amplitude vector solitons
in nonintegrable coupled nonlinear Schrödinger (NLS) equations was
studied in Ref. \citep{Yang_PRL00}, revealing a fractal structure
in the separation velocity versus collision velocity graph, explained
by a resonance mechanism between translational motion and internal
oscillations of the vector solitons. Also, the fractal structures
in soliton collisions were analyzed in Refs. \citep{Tan_PRE01,Zhu_PRE07,Zhu_PRL08,Teixeira_PLA16,Cardoso_ND19,Vergara_CSF21}.
The Gross-Pitaevskii equation with a repulsive delta function potential
was studied in Ref. \citep{Holmer_CMP07}. The interaction of composite
solitary waves, specifically breather collisions in a weakly discrete
sine-Gordon equation, was extensively studied in Refs. \citep{Dmitriev_PRE01,Dmitriev_PB02}.
In Ref. \citep{Dmitriev_PRE02}, the exact two-soliton solution to
the unperturbed NLS equation was analyzed, predicting strongly inelastic
soliton collisions with almost nonradiating characteristics, highly
sensitive to relative phase and independent of perturbation type.
In Ref. \citep{Craig_PF06}, the authors investigated pairwise nonlinear
interactions of solitary waves in water surface dynamics. In Ref.
\citep{Goodman_C08}, singular iterated maps, akin to Poincaré maps,
were derived to describe chaotic interactions between colliding solitary
waves, unveiling complex behavior including energy transfer to secondary
oscillation modes. Revisiting three-soliton collisions in the weakly
perturbed sine-Gordon equation, an effective three-particle model
was developed in Ref. \citep{Dmitriev_PRE08}. Soliton solutions of
two coupled nonlinear Schrödinger equations modulated in space and
time were presented in Refs. \citep{Cardoso_PLA10,Cardoso_PRE12}.
A numerical study of two-soliton collisions in a cubic nonlinear Schrödinger
equation system with small chaotic imperfections in the nonlinearity
was conducted in Ref. \citep{Cardoso_ND21}, revealing a direct relationship
between the Lyapunov coefficient and the formation of bonded/unbonded
soliton states. {The coded information storage pulsed
laser based on vector period-doubled pulsating solitons was designed
and implemented in Ref. \citep{Han_OLT23}.}

Recently, a study was conducted on soliton propagation in nonlinear
coupled waveguides described by coupled nonlinear Schrödinger equations.
This study introduced quasi-periodic nonlinear couplings to examine
their effect on soliton behavior, particularly within the exactly
integrable Manakov model framework \citep{Cardoso_BJP21}. Following
the previous study presented in Ref. \citep{Cardoso_BJP21}, this
work aims to investigate the influence of quasi-periodic coupling
on the dynamics of localized solutions in a system composed of three
coupled waveguides. {Indeed, understanding how different
modulation parameters influence soliton behavior is crucial for the
development of advanced technologies in optical communication and
signal processing. Additionally, investigating bound and unbound states
in such systems can offer new insights into the control and manipulation
of light pulses in optical fibers, contributing to the advancement
of high-speed communication networks and other photonic devices. In
this sense,} our study focuses on assessing the stability of the solutions
in the presence of quasi-periodic interaction, exploring the effects
of both the quasi-periodicity and perturbation amplitude.

The remainder of this article is organized as follows: In the next
section \ref{sec:2}, we provide a detailed description of the system
under study and the methodology employed in our analysis. Section
\ref{sec:3} presents the results of the numerical simulations, offering
an in-depth examination of the data and its implications. Finally,
conclusions drawn from our study are discussed in Section \ref{sec:4}.

\section{Theoretical Model}

\label{sec:2}

In this study, we conduct a comprehensive examination of a model based
on the interaction of three coupled optical fibers, characterized
by a medium with intrinsic cubic Kerr-type nonlinear properties that
permit solitonic solutions. This waveguide model is accurately described
by a set of NLS equations, which comprehensively capture the characteristics
and behaviors observed in the system under consideration. Thus, for
the pulse amplitude $\psi(z,t)$, we present the following NLS equations:
\begin{equation}
i\frac{\partial\psi_{i}}{\partial z}=-\frac{1}{2}\frac{\partial^{2}\psi_{i}}{\partial t^{2}}+\gamma\left(\left|\psi_{i}\right|^{2}+\sum^{3}_{\substack{j=1\\
j\neq i
}
}\delta_{ij}\left|\psi_{j}\right|^{2}\right)\psi_{i},\label{eq-geral}
\end{equation}
where $i=1,2,3$ and $\delta_{ij}=\delta_{ji}$. The parameter $\gamma$
represents the characteristic nonlinearity coefficient of the optical
fiber, and $\delta_{ij}$ denotes the cross-phase modulation coefficient,
responsible for the coupling between the pulse amplitudes {(for
more details on the design and theory of such a system, we refer readers
to Refs. \citep{Huang_JOSAA94,Akhmediev_JOSAB94,Gubeskys_EPJD04,Mumtaz_JLT13,Wright_NP22})}.
It is important to note that we obtain three independent and decoupled
NLS equations when $\delta_{ij}=0$. However, when $\delta_{ij}=1$,
the system transforms into a set of coupled equations with exact localized
solutions, analogous to the Manakov system for a model with three
fields. Specifically, when $\gamma<0$ and $\delta_{ij}=1$, corresponding
to the aforementioned Manakov system, the equations exhibit the following
localized solutions: 
\begin{align}
 & \psi_{1}=\frac{\sqrt{-\gamma}}{2}\operatorname{sech}\left[\frac{\gamma}{2}\left(t-t_{0}\right)\right]\cos\phi\sin\theta e^{i\sigma},\label{eq:psi1}\\
 & \psi_{2}=\frac{\sqrt{-\gamma}}{2}\operatorname{sech}\left[\frac{\gamma}{2}\left(t-t_{0}\right)\right]\sin\phi\sin\theta e^{i\eta},\label{eq:psi2}\\
 & \psi_{3}=\frac{\sqrt{-\gamma}}{2}\operatorname{sech}\left[\frac{\gamma}{2}\left(t-t_{0}\right)\right]\cos\theta,\label{eq:psi3}
\end{align}
with the amplitude of the solutions represented by $\theta$ and $\phi$,
with a phase difference between them given by $\sigma$ and $\eta$,
and the temporal position of the soliton center indicated by $t_{0}$.
Specifically, $\theta$ and $\phi$ quantify the peak amplitudes of
the solitary waves, while $\sigma$ and $\eta$ describe the relative
phase shifts between the interacting solitons. The parameter $t_{0}$
denotes the initial time at which the center of the soliton is located,
providing a reference point for tracking the soliton's temporal evolution.

One of the primary objectives of this work is to analyze the dynamics
exhibited by a system of coupled solitons, considering their coupling
determined by a quasi-periodic nonhomogeneous modulation coefficient.
In this context, our aim is to modify the cross-phase modulation coefficient
$\delta_{ij}$, represented by 
\begin{equation}
\delta_{ij}\equiv\delta(z)=\delta_{0}+\sum^{2}_{l=1}A_{l}\sin^{2}{(k_{l}z)},\label{delta_geral}
\end{equation}
where $\delta_{0}$ is a constant. Additionally, $A_{l}$ is associated
with the amplitude characterizing the non-linear coupling, while $k_{l}$
denotes the wave number. The dimensionless relationship between $k_{1}$
and $k_{2}$ is specifically defined as $k_{2}/k_{1}=(\sqrt{5}-1)/2$.
By incorporating these quasi-periodic modifications to the cross-phase
modulation coefficient, we intend to investigate the impact on the
stability and dynamics of the coupled soliton solutions. This investigation
includes detailed numerical simulations and theoretical analyses aimed
at understand how the quasi-periodic nature of $\delta(z)$ influences
soliton interactions, propagation characteristics, and potential applications
in nonlinear optics. The inclusion of quasi-periodic terms allows
us to explore more complex and realistic scenarios, providing deeper
insights into the behavior of coupled soliton systems under quasi-periodic
perturbations. Our study will also examine the resulting soliton behavior
under various parameter settings for $A_{l}$ and $k_{l}$, assessing
the robustness of soliton solutions and identifying conditions that
lead to stable or unstable dynamics.

{To realize quasiperiodic nonhomogeneous modulation
in optical systems, several approaches can be employed. One can use
optical waveguides with tunable materials, such as photoelastic or
liquid crystal materials, to achieve spatially varying refractive
indices \citep{Smalyukh_OE07}. Electro-optic modulators can introduce
quasiperiodic phase shifts or amplitude modulations \citep{Xu_NAT20}.
Fabrication techniques like photonic crystals, diffraction gratings,
or fiber Bragg gratings can create quasiperiodic patterns in waveguides
\citep{Hill_APL93}. Surface acoustic waves can modulate the refractive
index in fibers, while adaptive optics systems can provide real-time
control of modulation \citep{Poveda_JPD19}. Hybrid approaches combining
these methods offer precise control over quasiperiodic modulation,
enabling detailed experimental studies of soliton dynamics and other
phenomena in complex media.}

\section{Numerical Results}

\label{sec:3}

The numerical procedure employed in this study is based on the split-step
method, which is a numerical technique for solving nonlinear differential
equations, particularly the NLS equation. This method divides each
step into two or more parts: one for the linear part and the other
for the nonlinear part. We combine this technique with the Fast Fourier
Transform (FFT) method, known for its ability to significantly reduce
the computational time required for processing compared to direct
methods, thus providing greater efficiency in numerical resolution.
This combined approach allows for accurate and computationally efficient
simulations of the dynamics of coupled solitons, enabling detailed
analysis of their behavior under various conditions \citep{Vesely01,Yang10}.

In order to analyze the dynamics of soliton propagation in coupled
waveguides, we will investigate two distinct cases for our model.
The first case involves using as input profiles the solutions for
the scenario analogous to the Manakov model, with $\gamma=-1$ and
initially $\delta(z=0)=1$, which we refer to as correlated solutions.
The second case pertains to solutions with their peaks initially localized
at different independent time instants, presenting individual behaviors
without interactions between them when $z=0$. By examining these
two cases, our aim is to elucidate the influence of initial conditions
on the propagation dynamics of solitons in coupled waveguide systems.

\subsection{Correlated Solutions}

\label{subsec:3.1}

Our first case under analysis consists of the correlated system configuration
mentioned above, Eqs. \eqref{eq:psi1}-\eqref{eq:psi3}, with $\delta(z=0)=1$
and $\gamma=-1$. However, we aim to induce a breakdown in the integrability
of the system by making alterations to the cross-phase modulation
coefficient. To investigate the effects of quasi-periodic modulation,
as defined in Eq. \eqref{delta_geral}, we set the values of $A_{l}$
as $A_{1}=A_{0}/2$, $A_{2}=A_{0}/2$, and $\delta_{0}=1$. We will
then consider 
\begin{equation}
\delta(z)=1+\frac{A_{0}}{2}\left[\sin^{2}(k_{1}z)+\sin^{2}(k_{2}z)\right],\label{deltaperiodico}
\end{equation}
with amplitude of the quasi-periodic perturbation denoted by $A_{0}$,
where $k_{1}$ and $k_{2}$ obey the dimensionless ratio given by
$k_{2}/k_{1}=(\sqrt{5}-1)/2$, as previously mentioned. The amplitudes
of the solutions of Eqs. \eqref{eq:psi1}-\eqref{eq:psi3} are determined
by the angles $\theta$ and $\phi$, such that, for the correlated
case, all three solitons have the same initial amplitude 
\begin{equation}
\theta=\cot^{-1}\left(\frac{\sqrt{2}}{2}\right)\text{ \, and \, }\phi=\frac{\pi}{4}.
\end{equation}

\begin{figure}[tb]
\includegraphics[width=1\columnwidth]{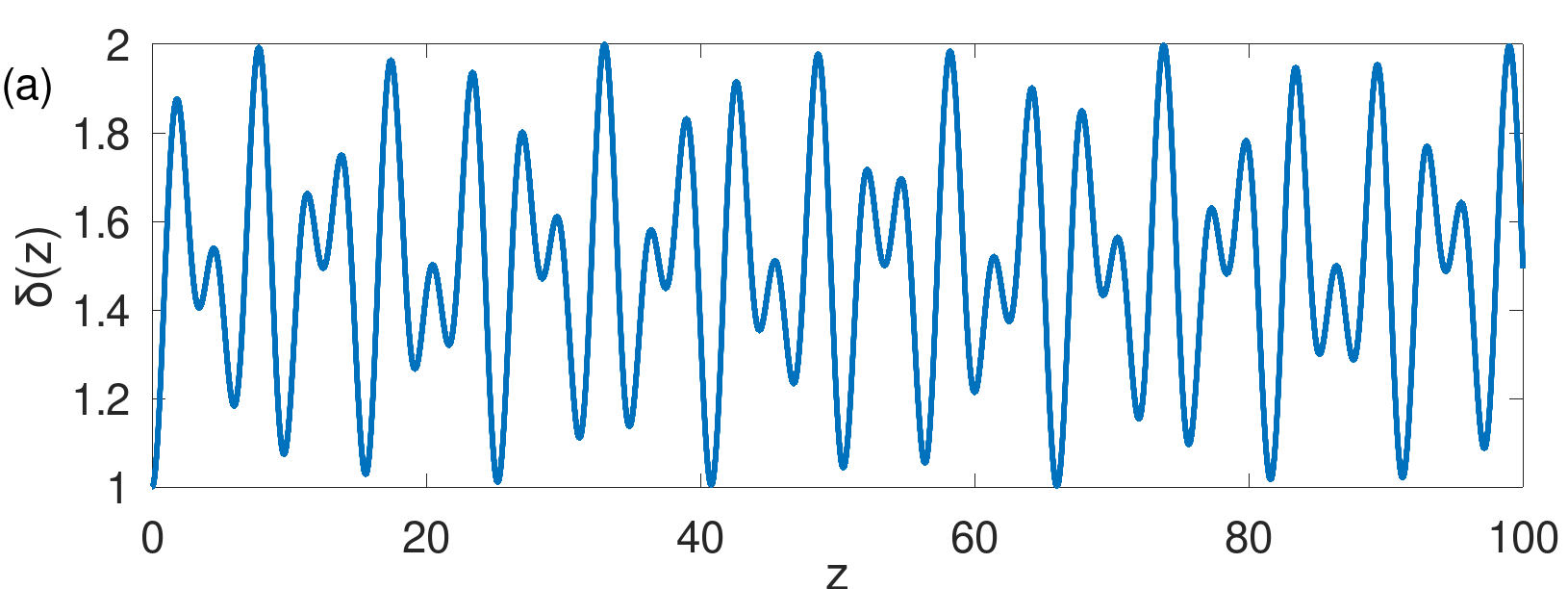} 
\includegraphics[width=1\columnwidth]{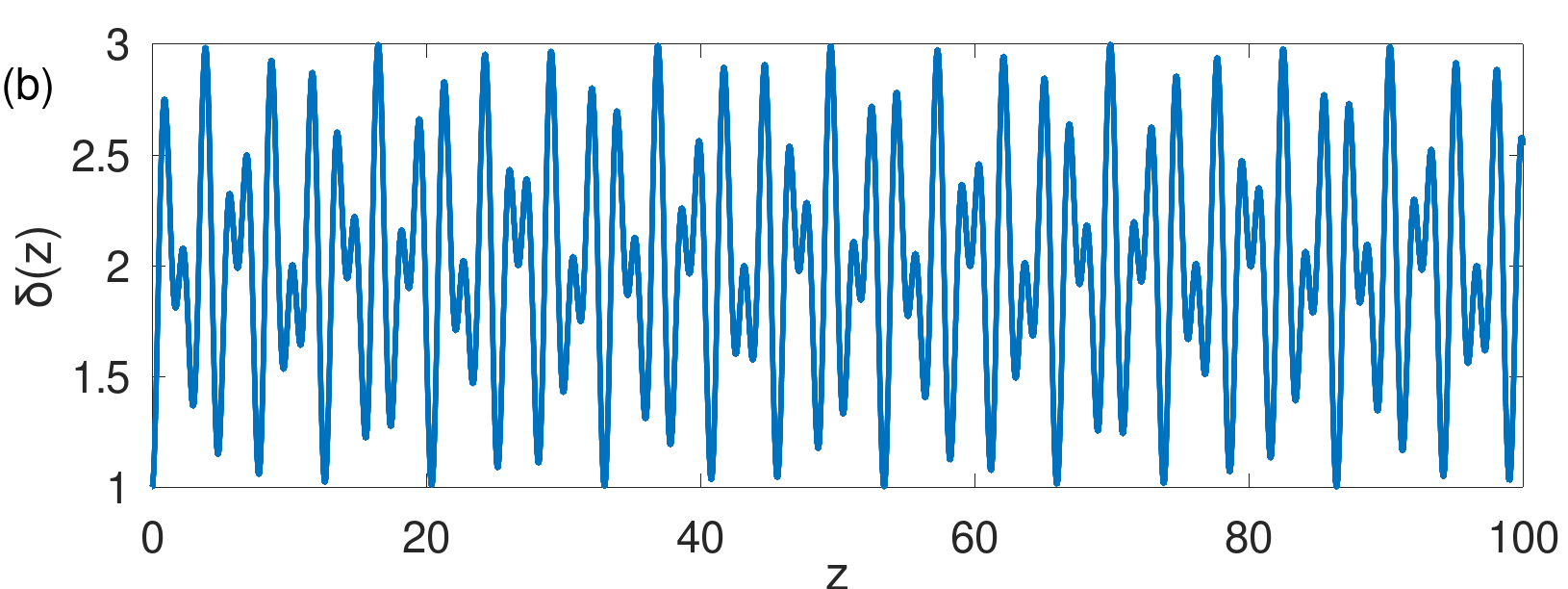}
\caption{Quasi-periodic cross-phase modulation coefficient $\delta(z)$, given
by Eq. \eqref{deltaperiodico}. In panel (a), we depict the modulation
for $k_{1}=1$ and $A_{0}=1$, while in panel (b), we show $k_{1}=2$
with $A_{0}=2$. For both cases, $k_{2}$ follows the ratio $k_{2}/k_{1}=(\sqrt{5}-1)/2$.}
\label{Fig.1-delta} 
\end{figure}

In Fig. \ref{Fig.1-delta}, we depict the profile of $\delta(z)$
given by Eq. \eqref{deltaperiodico} and illustrate the influence
of $k_{1}$ and $A_{0}$ on the amplitude of interactions between
the fields. It can be observed that variations in these coefficients
have the potential to amplify the level of disorder in the system,
directly impacting in the stability and behavior of the interactions.
With this perturbation, we will be able to analyze the resulting effects
introduced into the system, and consequently, their influence on the
system dynamics.

\begin{figure}[tb]
\includegraphics[width=1\columnwidth]{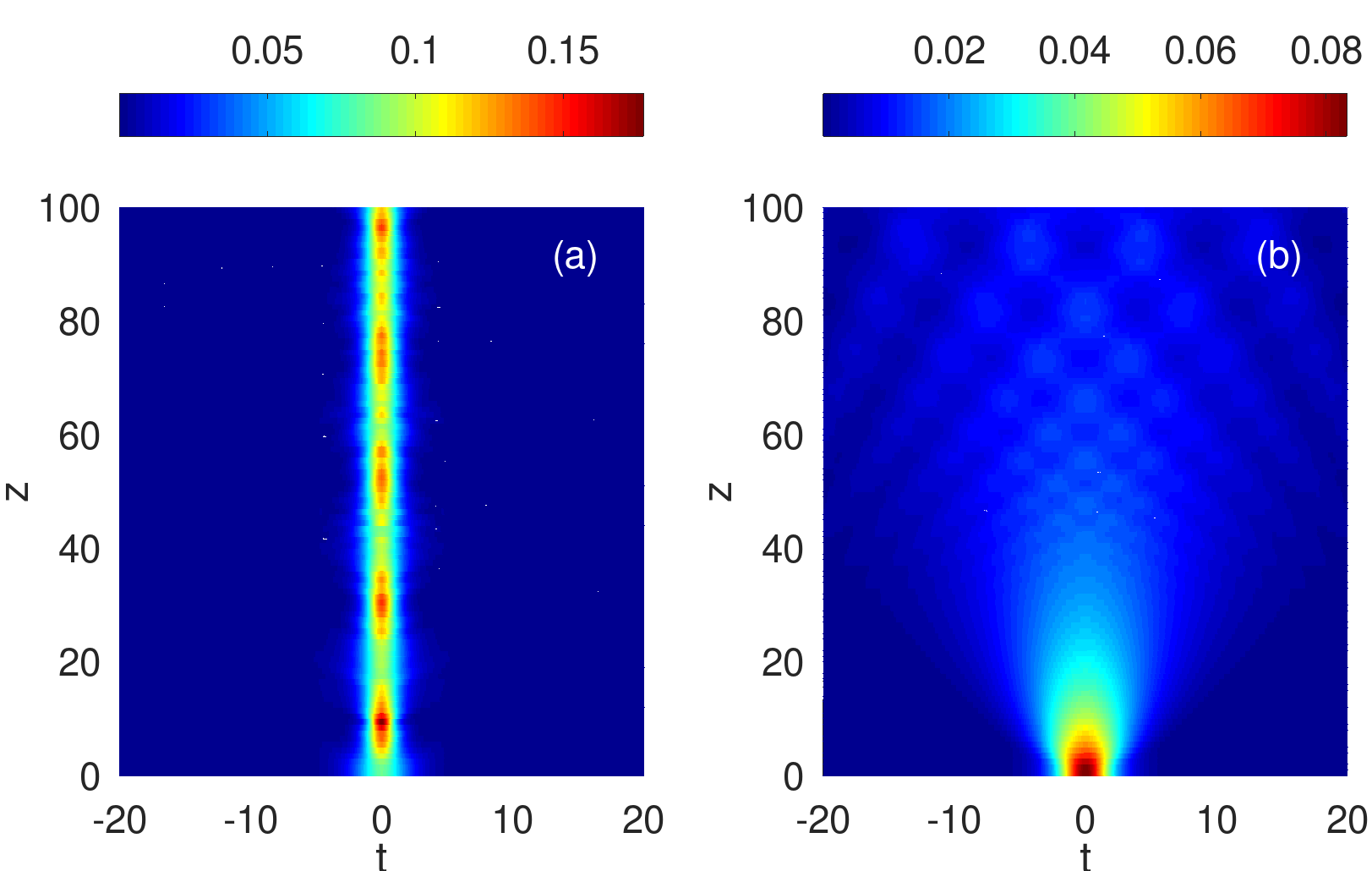}
\caption{Spatial evolution analysis of the profile of $\left|\psi_{1}\right|^{2}$
for the cases with a) $A_{0}=2$ and b) $A_{0}=-2$. In both cases,
we use $\gamma=-1$, $k_{1}=1$, and $k_{2}=(\sqrt{5}-1)/2$. The
color bar indicates the intensity of $\left|\psi_{1}\right|^{2}$.}
\label{Fig.2-evol-corr} 
\end{figure}

In Fig. \ref{Fig.2-evol-corr}, we illustrate an example of the instability
generated by the perturbation in the cross-phase modulation term,
showing the evolution of the profile of $\left|\psi_{1}\right|^{2}$
in two distinct cases with different intensities for the amplitude
$A_{0}$. In Fig. \ref{Fig.2-evol-corr}(a), where $A_{0}=2$, we
observe a stable solution, albeit with oscillations in the peak intensity.
Meanwhile, in Fig. \ref{Fig.2-evol-corr}(b), for $A_{0}=-2$, we
demonstrate an unstable solution, characterized by a significant decrease
in amplitude, leading to pulse dissipation into radiation. This example
clearly demonstrates how variations in the perturbation amplitude
can significantly affect the behavior and stability of the system,
as reflected in the distribution of the field $\left|\psi_{1}\right|^{2}$
along the fiber. Similar behavior is observed for the fields $\psi_{2}$
and $\psi_{3}$.

\begin{figure}[tb]
\centering \includegraphics[width=1\columnwidth]{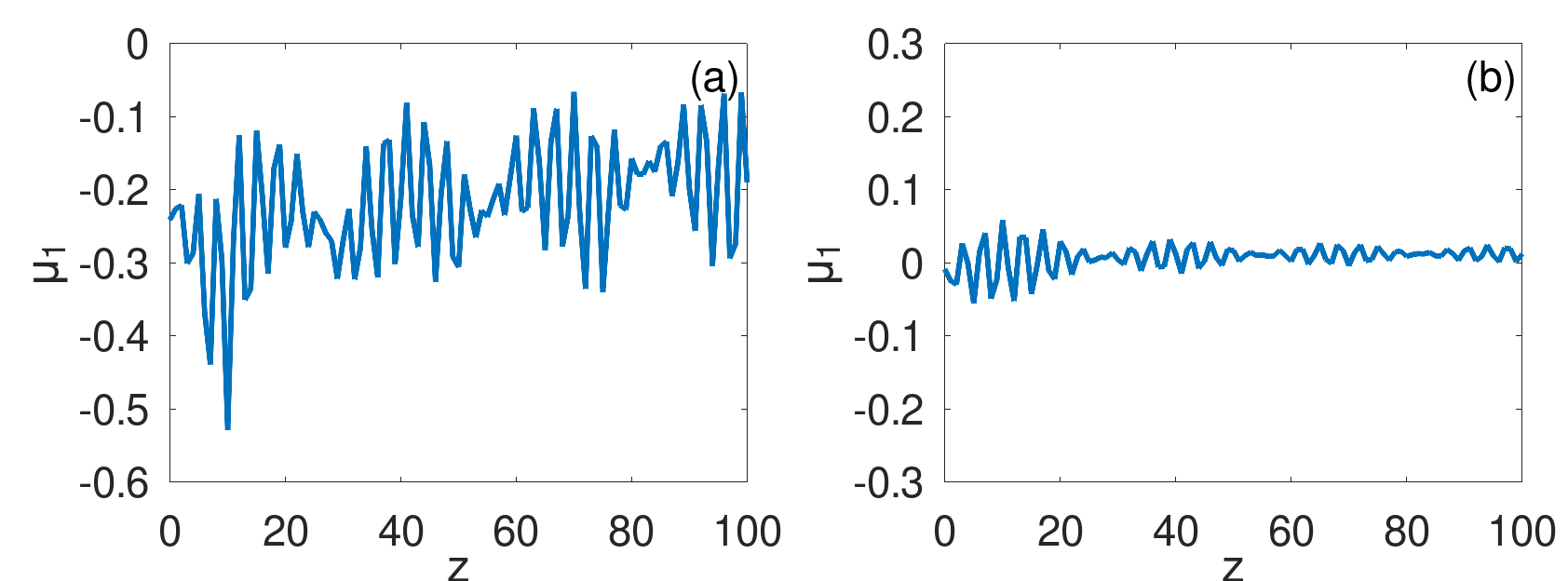}
\includegraphics[width=1\columnwidth]{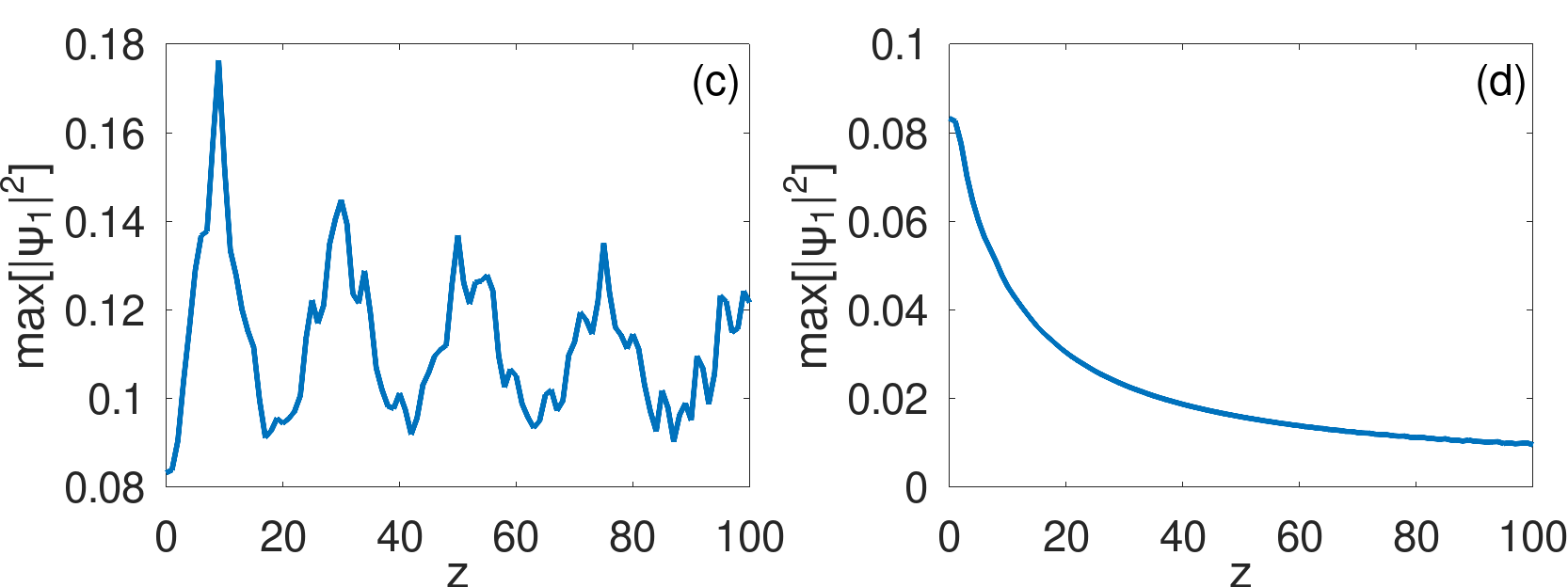} 
\caption{Upper panel: Propagation constant $\mu_{1}$ for the cases with a)
$A_{0}=2$ and b) $A_{0}=-2$. Lower panel: Peak intensity of the
solution $|\psi_{1}|^{2}$ for the cases with c) $A_{0}=2$ and d)
$A_{0}=-2$.}
\label{Fig.3-a.mu-b.pico} 
\end{figure}

Additionally, we conducted calculations for the maximum peak intensity
and the propagation constant $\mu_{i}$. The propagation constant
was obtained by substituting $\psi_{i}(t,z)=\psi_{i}(t)e^{-i\mu_{i}z}$
into Eqs. \eqref{eq:psi1}-\eqref{eq:psi3}. This substitution results,
for example, in the following equation for the field $\psi_{1}$:
\begin{equation}
\begin{split}\mu_{1}= & \frac{1}{\int|\psi_{1}|^{2}dt}\left[-\frac{1}{2}\int\frac{\partial^{2}}{\partial t^{2}}|\psi_{1}|^{2}dt+\gamma\int|\psi_{1}|^{4}dt+\right.\\
+ & \left.\gamma\delta(z)\int{|\psi_{2}|^{2}|\psi_{1}|^{2}}dt+\gamma\delta(z)\int{|\psi_{3}|^{2}||\psi_{1}|^{2}}dt\right],
\end{split}
\label{eq-mu1}
\end{equation}
where the integration is performed over the entire temporal domain.
The results of these analyses are presented in the plots of Fig. \ref{Fig.3-a.mu-b.pico},
providing a visual representation of the variations of these parameters
along the spatial direction. In Fig. \ref{Fig.3-a.mu-b.pico}(a) and
\ref{Fig.3-a.mu-b.pico}(b), we can compare the fluctuation experienced
by the propagation constant for the cases where $A_{0}=2$ and $A_{0}=-2$,
respectively. We observe a noticeable decrease in magnitude and a
tendency to oscillate around zero for the case where $A_{0}=-2$,
indicating that the solution becomes unstable for this value of the
quasi-periodic perturbation amplitude.

Furthermore, in Fig. \ref{Fig.3-a.mu-b.pico}(c), we plot the peak
of the solution for the case where $A_{0}=2$, showing an initial
increase in pulse height followed by oscillations resembling a resonance
window. Simultaneously, in Fig. \ref{Fig.3-a.mu-b.pico}(d), a significant
decrease in the solution peak is observed for the case where $A_{0}=-2$,
indicating pulse dissipation along the optical fiber. The results
obtained, demonstrated here for $\mu_{1}$ and for the peak of the
field $\psi_{1}$, are identical for the fields $\psi_{2}$ and $\psi_{3}$,
owing to the initial correlation in our system.

\subsection{Uncorrelated Solutions}

\label{sec:subthree2}

For the second case under analysis, we explored localized, independent
and uncorrelated solutions. These solutions were derived by considering
the quasi-periodic perturbation term such that $\delta(z=0)=0$. In
this specific configuration, the decoupled solutions of the fields
are expressed as 
\begin{align}
\psi_{i}(t,z)=\frac{\sqrt{-\gamma}}{2}\operatorname{sech}\left[\frac{\gamma}{2}(t-t_{i})\right]e^{-i\mu_{i}z},\label{eq:psi-geral-N-Corr}
\end{align}
with $i=1,2,3$ and $t_{i}$ representing the arbitrary positions
for the peak of the solutions of the respective fields $\psi_{i}$.
To analyze the effects of the quasi-periodic modulation on the interaction
between the solitons, we chose a symmetric scenario for the initial
peak positions of the solutions: $t_{3}=-t_{1}=t_{0}$; furthermore,
for the second field, we set $t_{2}=0$. Thus, the soliton solutions
are temporally separated depending on $t_{0}$. This initial configuration
aims to provide a precise comparative analysis of the effects of quasi-periodic
modulation on the interaction and behavior of the solitons along the
spatial domain.

To satisfy the uncorrelated configuration, we set $\delta_{0}=0$
for the quasi-periodic cross-phase modulation coefficient, as given
by Eq. \eqref{delta_geral}, keeping the values of $A_{l}$ as $A_{1}=A_{0}/2$
and $A_{2}=A_{0}/2$. Consequently, the expression for the quasi-periodic
modulation coefficient becomes 
\begin{align}
\delta(z)=\frac{1}{2}A_{0}\left[\sin^{2}(k_{1}z)+\sin^{2}(k_{2}z)\right],\label{delta_uncorrelated}
\end{align}
where the ratio $k_{2}/k_{1}=(\sqrt{5}-1)/2$ is maintained. In this
case, for $A_{0}=0$, we do not revert to the solutions analogous
to the Manakov system, as expressed by Eq. \eqref{eq-geral}. Instead,
when $A_{0}=0$, we obtain three independent NLS equations, and their
solutions are given by Eq. \eqref{eq:psi-geral-N-Corr}. This highlights
the distinct behavior of the system under different quasi-periodic
modulation amplitudes, where $A_{0}=0$ leads to a decoupled scenario,
providing insight into the nature of soliton interactions without
cross-phase modulation effects.

\begin{figure}
\centering \includegraphics[width=1\columnwidth]{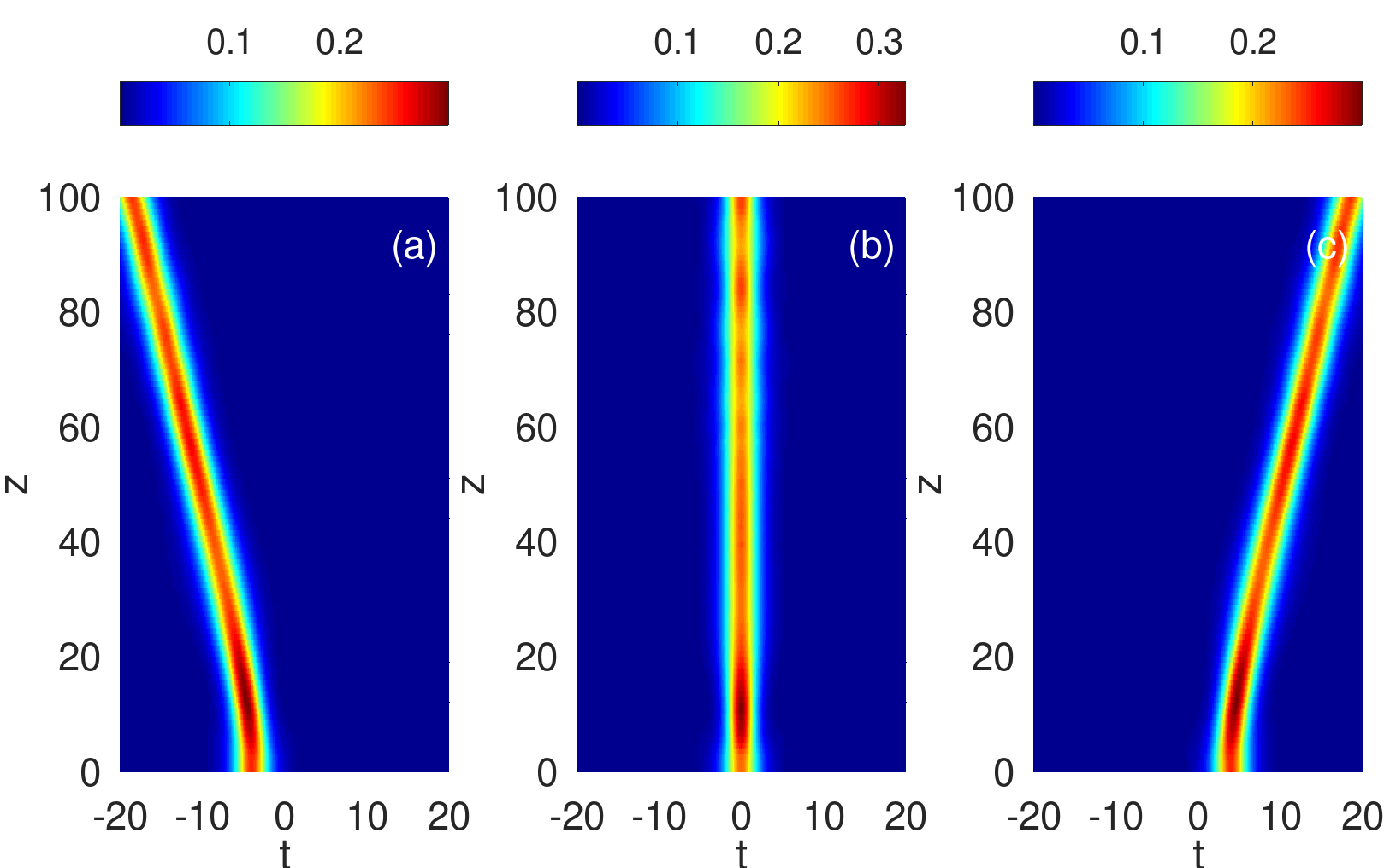}
\caption{Spatial evolution of the field profiles: a) $|\psi_{1}|^{2}$, b)
$|\psi_{2}|^{2}$, and c) $|\psi_{3}|^{2}$ with $k_{1}=1$ and $A_{0}=-0.8$,
highlighting a repulsive interaction leading to continuous separation
of solitons.}
\label{Fig.4-n-corr.A0-08} 
\end{figure}

The influence of quasi-periodic modulation is explored by considering
different values for the perturbation amplitude $A_{0}$, including
both positive and negative values. When $A_{0}<0$, a repulsive interaction
occurs, causing a continuous separation of the solitons, as there
is no confining effect in the system. Using $A_{0}=-0.8$, we plot
the evolution of the fields $\psi_{1}$, $\psi_{2}$, and $\psi_{3}$
in Figs. \ref{Fig.4-n-corr.A0-08}(a), \ref{Fig.4-n-corr.A0-08}(b),
and \ref{Fig.4-n-corr.A0-08}(c), respectively. If the solitons were
initially positioned at the same time instant, the repulsion between
the fields would result in a decay of the solution, similar to what
was observed in the correlated case. However, by positioning the solutions
at distinct time instants, the repulsive effect causes a continuous
separation between the solitons. This dynamic reflects the direct
influence of the repulsive interaction generated by the quasi-periodic
modulation, leading to the solitons separation when initially placed
at different temporal offsets.

\begin{figure}
\centering \includegraphics[width=1\columnwidth]{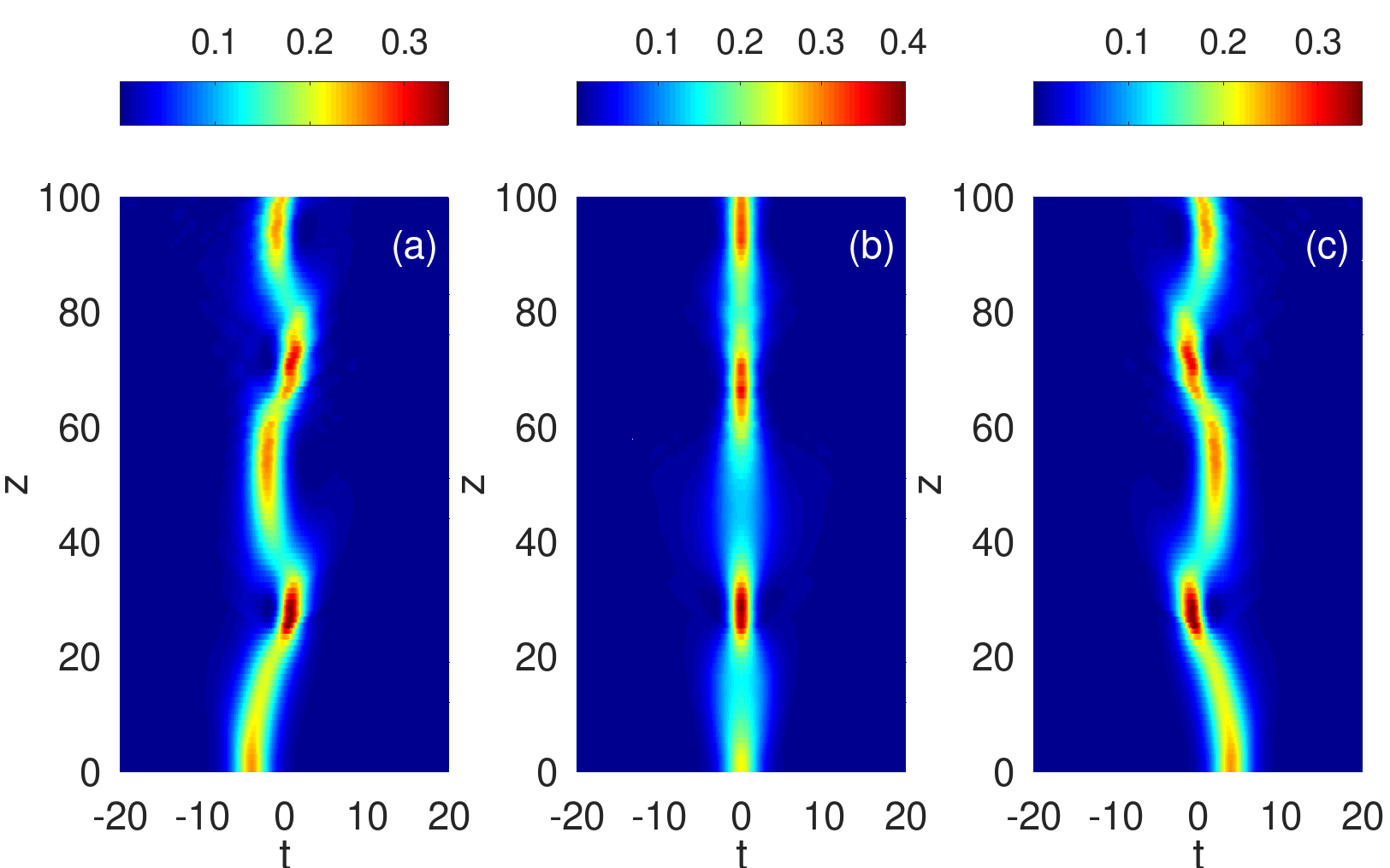}
\caption{Spatial evolution of the field profiles: a) $|\psi_{1}|^{2}$, b)
$|\psi_{2}|^{2}$, and c) $|\psi_{3}|^{2}$ with $k_{1}=1$ and $A_{0}=0.8$,
indicating an attraction between the solitons.}
\label{Fig.5-n-corr.A0+08} 
\end{figure}

When $A_{0}>0$, we observe an attractive interaction between the
fields. Setting $A_{0}=0.8$, we plotted the evolution of the peak
amplitudes of the fields $\psi_{1}$, $\psi_{2}$, and $\psi_{3}$,
depicted in Fig. \ref{Fig.5-n-corr.A0+08}(a), Fig. \ref{Fig.5-n-corr.A0+08}(b),
and Fig. \ref{Fig.5-n-corr.A0+08}(c), respectively. As the parameter
$A_{0}$ increases, the attraction between the fields intensifies.
{Indeed, an increase in $A_{0}$ results in an increase
in the average value of $\delta(x)$ in both cases considered here,
as can be directly observed from Eqs. (\ref{deltaperiodico}) and
(\ref{delta_uncorrelated}).} Specifically, the solitons begin to
converge and interact more strongly due to the increased attractive
force. This behavior indicates the significant role of quasi-periodic
modulations in influencing the solitons dynamics, where modulation
with a positive amplitude enhances the coupling between the solitons,
promoting a coherent and synchronized interaction pattern.

\begin{figure}[tb]
\centering \includegraphics[width=1\columnwidth]{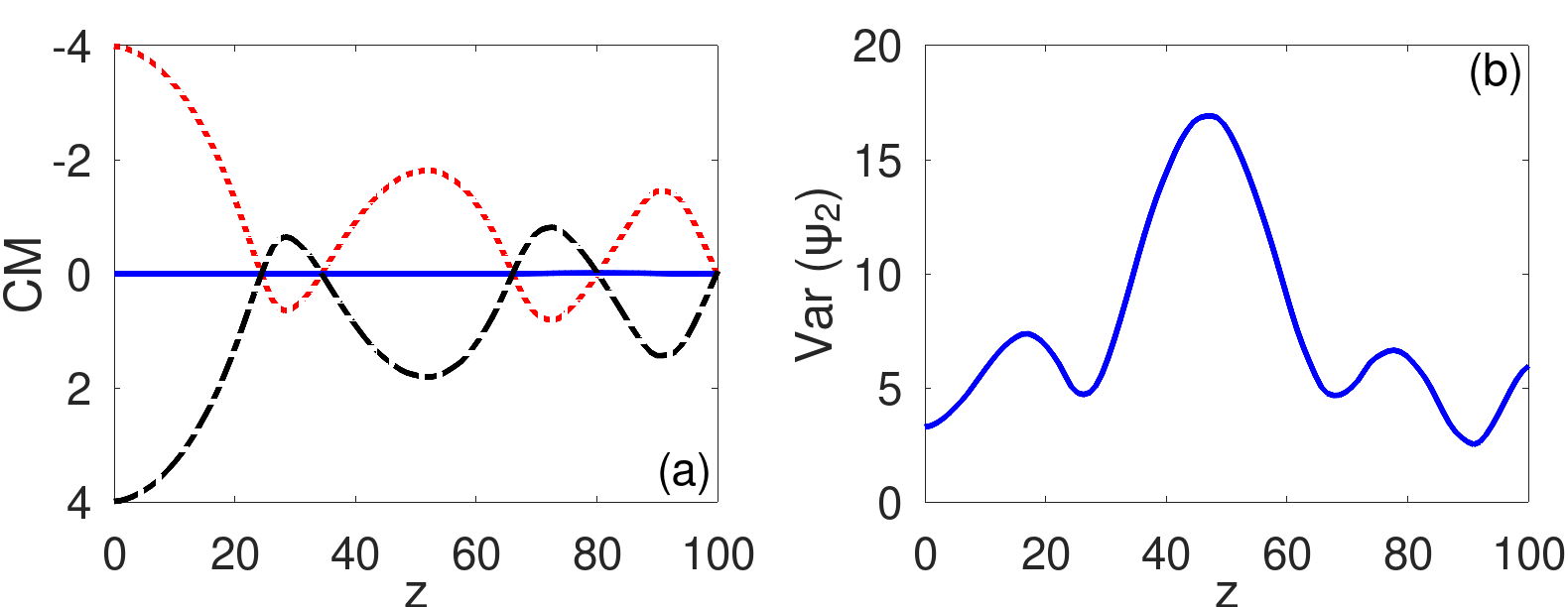}
\caption{In panel (a), the center of mass of the solutions for the fields $\psi_{1}$,
$\psi_{2}$, and $\psi_{3}$ are represented by dotted red, solid
blue and dashed black curves, respectively. In panel (b), the variance
of $\psi_{2}$ is depicted, providing insight into the spread and
dispersion of the field. The parameters used are $A_{0}=0.8$ and
$k_{1}=1$.}
\label{Fig.6-a.CM-b.VAR} 
\end{figure}

In Fig. \ref{Fig.6-a.CM-b.VAR}, we present the plots depicting the
center of mass and the variance of the solutions. Fig. \ref{Fig.6-a.CM-b.VAR}(a)
shows the center of mass of the solutions, where $\psi_{1}$, $\psi_{2}$,
and $\psi_{3}$ are represented by dotted red, solid blue, and dashed
black curves, respectively. These plots indicate the presence of a
bound state for these interaction values between the fields. In Fig.
\ref{Fig.6-a.CM-b.VAR}(b), the variance of the field $\psi_{2}$
is plotted, defined by 
\begin{align}
\text{Var}(\psi_{i})=\langle\psi^{2}_{i}\rangle-\langle\psi_{i}\rangle^{2},
\end{align}
where $\langle\cdot\rangle$ denotes the average over the temporal
domain. The variance plot provide insight into the spread and stability
of the soliton solution. The center of mass plot shown in Fig. \ref{Fig.6-a.CM-b.VAR}(a)
corresponds to a three-field bound state and indicates an oscillatory
propagation of these solitons. Conversely, deviations in variance
can highlight the onset of instability and increased dispersion within
the system. In fact, interaction between the fields will cause a change
in the value of the variance. In the observed bound state, for the
field $\psi_{2}$ (Fig. \ref{Fig.6-a.CM-b.VAR}(b)), the variance
decreases when the solitons approach each other and increases when
they move apart. These detailed analyses of the center of mass and
variance elucidate the intricate dynamics of the coupled soliton system
under the influence of quasi-periodic modulation.

\subsubsection*{Influence of the Modulation Amplitude}

Now, we investigate the impact of the modulation amplitude on the
dynamics of the coupled soliton system. By systematically varying
the amplitude of the quasi-periodic modulation term, denoted as $A_{0}$,
we aim to uncover the resultant changes in the behavior and stability
of the solitons.

By significantly increasing the amplitude $A_{0}$ of the quasi-periodic
oscillation, we observe a breakdown in the symmetry of the bound solutions
in the system. This phenomenon triggers the formation of unbound states,
where the solutions continuously move away from each other without
a defined distance limit.

This induced symmetry breakdown can be observed by comparing two distinct
scenarios. In the first scenario, we consider the interaction between
the fields based solely on the constant $\delta_{0}$ of the cross-phase
modulation coefficient Eq. \eqref{delta_geral}, i.e., $\delta(z)=\delta_{0}$,
with $A_{0}=0$. In the second case, we set $\delta_{0}=0$, considering
the interaction given only by the second term of Eq. \eqref{delta_geral},
responsible for the quasi-periodicity of the interactions between
the fields. Thus, we have $\delta(z)=A_{0}/2(\sin^{2}(k_{1}z)+\sin^{2}(k_{2}z))$.
Taking into account the properties of the average of $\sin^{2}$ and
also observed in Fig. \ref{Fig.1-delta}, we compare the solutions
given by $\delta_{0}$ and $2A_{0}$. Essentially, this comparison
contrasts the constant interaction between the fields with the averaged
interaction arising exclusively from quasi-periodic modulation.

\begin{figure}[tb]
\centering \includegraphics[width=1\columnwidth]{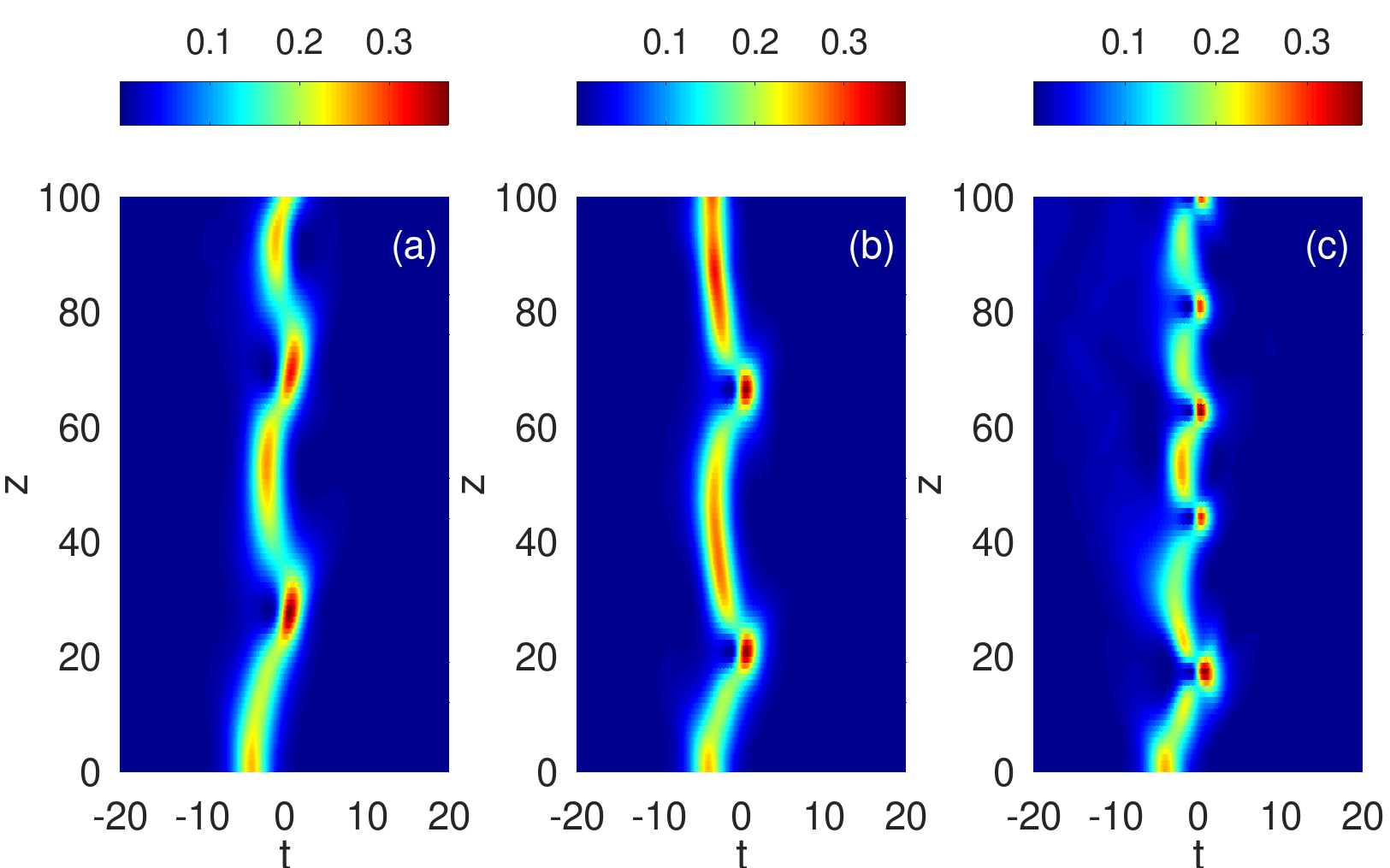}
\includegraphics[width=1\columnwidth]{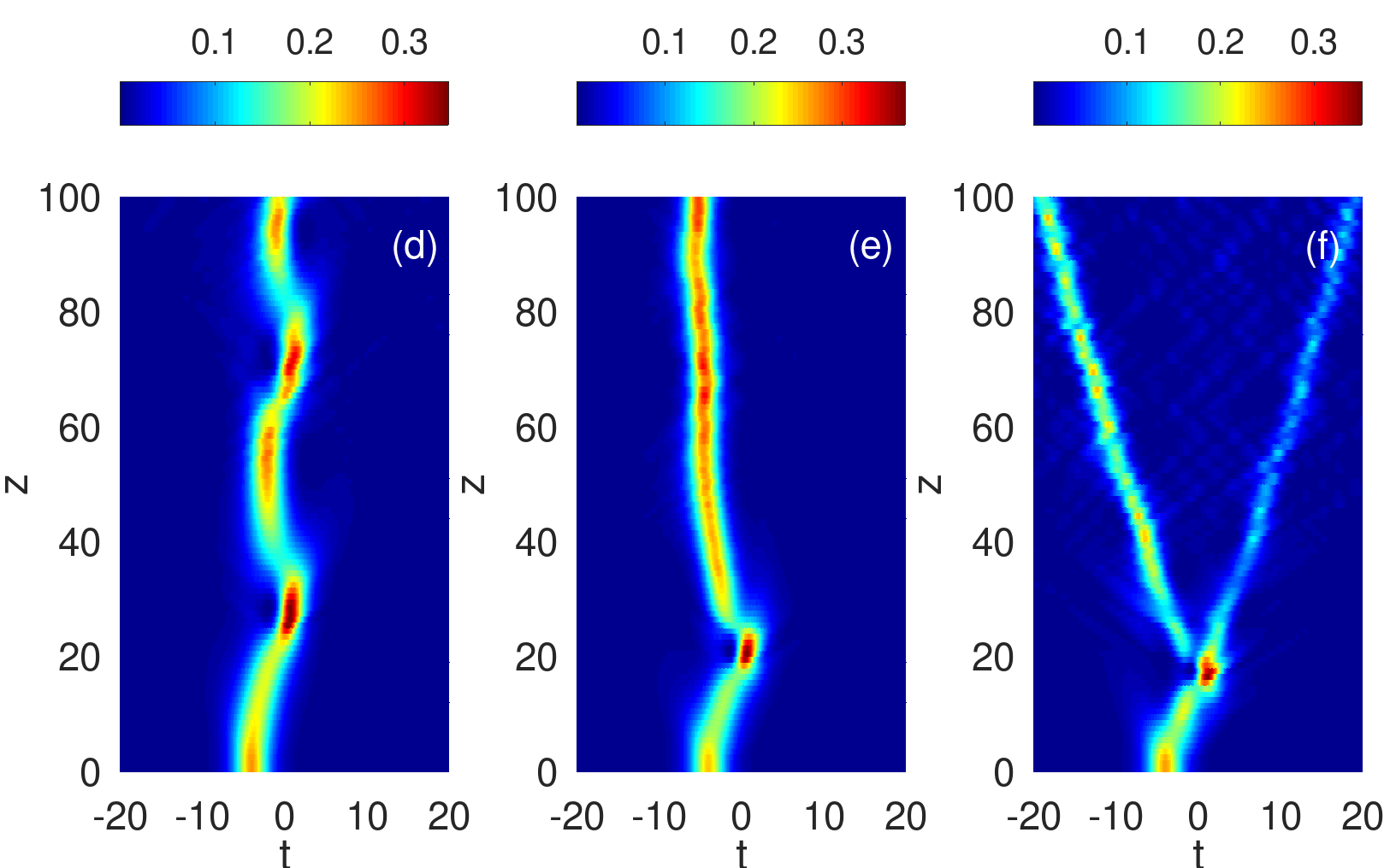}
\caption{Comparison of the evolution of the profile of $|\psi_{1}|^{2}$ with
constant interaction $\delta_{0}$ (upper panel with $A_{0}=0$) and
quasi-periodic interaction (lower panel with $\delta_{0}=0$). In
panel a), $\delta_{0}=0.4$; in panel b), $\delta_{0}=0.8$; and in
panel c), $\delta_{0}=1.5$. For the quasi-periodic interaction, for
comparison purposes, in panel d) we have $A_{0}=0.8$; in panel e),
$A_{0}=1.6$; and in panel f), $A_{0}=3.0$.}
\label{Fig.7-up.const-down.periodic} 
\end{figure}

In Fig. \ref{Fig.7-up.const-down.periodic}, profiles of $|\psi_{1}|^{2}$
for two distinct scenarios are presented: one with constant interaction
(upper panel) and the other with exclusively quasi-periodic interaction
(lower panel). For comparison purposes, pairs of values for the parameters
$\delta_{0}$ and $A_{0}$ are used. In Figs. \ref{Fig.7-up.const-down.periodic}(a)-\ref{Fig.7-up.const-down.periodic}(c),
we plot the cases with $\delta_{0}=0.4$, $0.8$, and $1.5$, respectively,
all of them with $A_{0}=0$. These panels illustrate the effects of
varying the constant interaction strength $\delta_{0}$ while keeping
quasi-periodic modulation absent ($A_{0}=0$). In Figs. \ref{Fig.7-up.const-down.periodic}(d)-\ref{Fig.7-up.const-down.periodic}(f),
we fix $\delta_{0}=0$ and vary $A_{0}=0.8$, $1.6$, and $3.0$,
respectively. Here, the panels demonstrate the influence of increasing
quasi-periodic modulation amplitude $A_{0}$ while keeping the constant
interaction ($\delta_{0}=0$) negligible.

A striking similarity in the evolution of the solution is observed
for low values of interaction between the fields. However, as we increase
the interaction, characterized by $\delta_{0}=0.8$ and $A_{0}=0$
in Fig. \ref{Fig.7-up.const-down.periodic}(b), and $A_{0}=1.6$ with
$\delta_{0}=0$ in Fig. \ref{Fig.7-up.const-down.periodic}(e), a
divergence is evident. Specifically, there is a slight delay in the
trajectory of the field $\psi_{1}$ when quasi-periodic interaction
is present.

This discrepancy becomes more pronounced in the latter cases presented:
$\delta_{0}=1.5$ with $A_{0}=0$ (Fig. \ref{Fig.7-up.const-down.periodic}(c)),
and $A_{0}=3.0$ with $\delta_{0}=0$ (Fig. \ref{Fig.7-up.const-down.periodic}(f)).
In these scenarios, it is notable that for sufficiently high values
of quasi-periodic modulation, the solitons evolution in the optical
fiber deviates significantly from the expected behavior with constant
interaction.

In Fig. \ref{Fig.7-up.const-down.periodic}(f), particularly, a scattering-like
pattern can be identified, evidenced after a brief interaction period
around $z=20$. This result reinforces the substantial influence of
quasi-periodic modulation on the dynamics and behavior of solitons
along the optical fiber.

\subsubsection*{Influence of Wave Vector}

Interestingly, the solutions of the system with exclusively quasi-periodic
interaction return to a bound state when the wave vector $k_{1}$
is increased. This behavior becomes evident for values of $k_{1}$
greater than a threshold around $k_{1}\approx2$. This phenomenon
highlights the system's sensitivity to variations in the wave vector
$k_{1}$, where an increase in this parameter results in the reintegration
of solutions into a bound state following their dissociation induced
by the high amplitude of the quasi-periodic oscillations.

\begin{figure}[tb]
\centering \includegraphics[width=1\columnwidth]{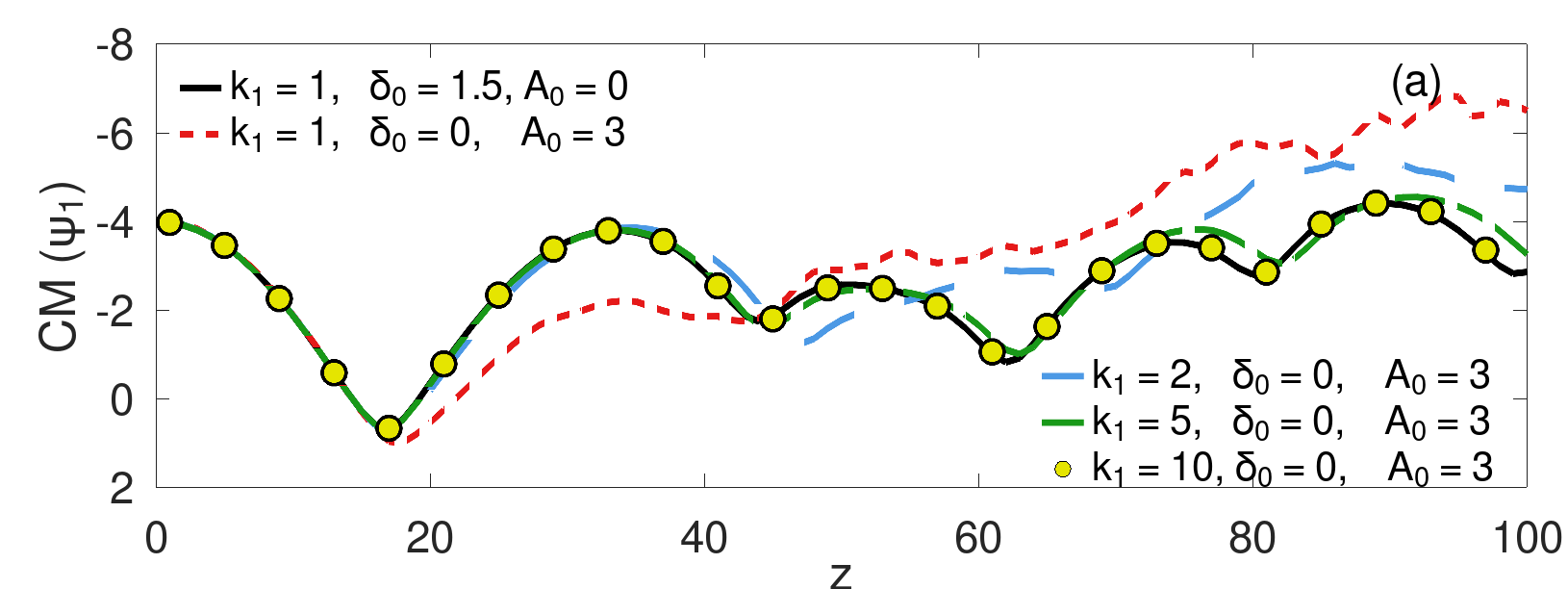}
\includegraphics[width=1\columnwidth]{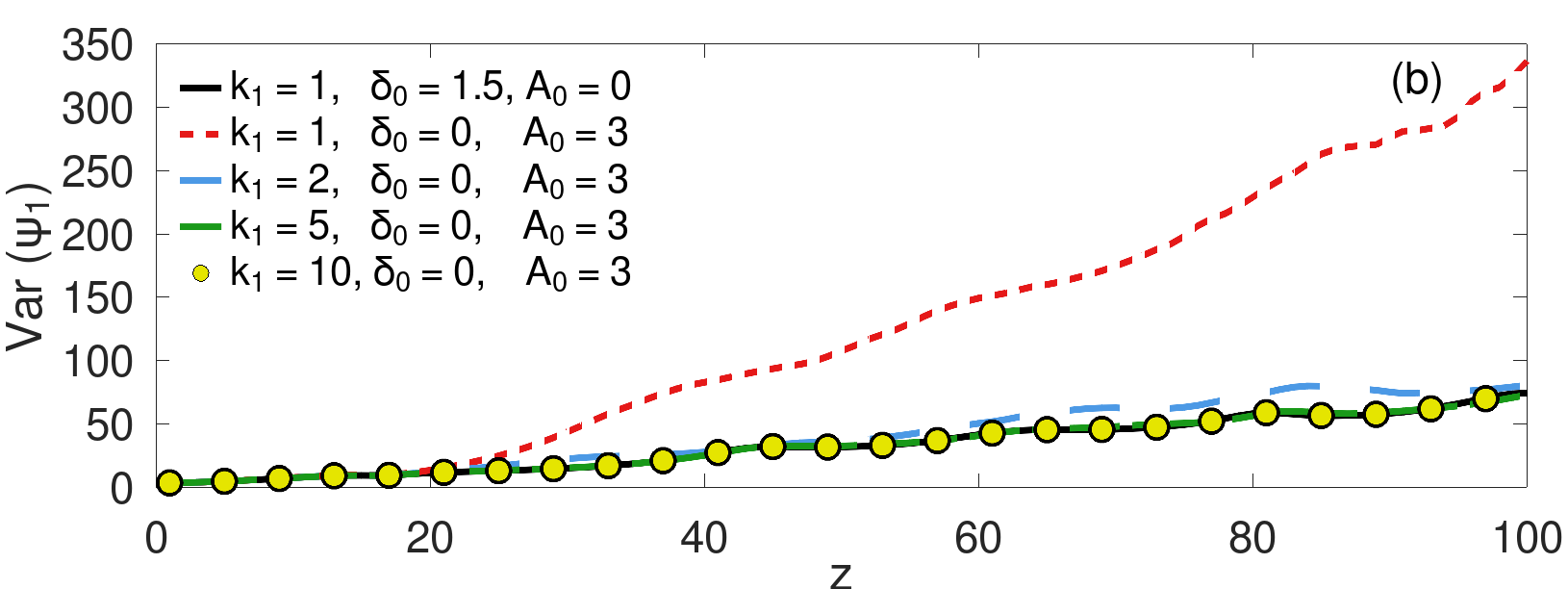} \caption{(a) Center of mass and (b) variance of $\psi_{1}$ for two distinct
systems. In the solid black line, the interaction between the fields
is constant, given by $\delta_{0}=1.5$, with $k_{1}=1$ and $A_{0}=0$.
In the other curves, we have $\delta_{0}=0$ and $A_{0}=3$ with $k_{1}=1$
(red dotted line), $k_{1}=2$ (blue dashed line), $k_{1}=5$ (green
dash-dotted line), and $k_{1}=10$ (yellow circles).}
\label{Fig.8-a.CM-b.VAR-k1-maior} 
\end{figure}

In Fig. \ref{Fig.8-a.CM-b.VAR-k1-maior}, we present the results of
the center of mass and variance of the solution for $\psi_{1}$. Fig.
\ref{Fig.8-a.CM-b.VAR-k1-maior}(a), shows the solid black line, illustrating
the center of mass of $\psi_{1}$ under exclusive constant interaction,
represented by $\delta_{0}=1.5$, with $A_{0}=0$ and $k_{1}=1$.
The other curves depict different configurations where the interaction
is governed solely by quasi-periodic modulation. For instance, the
dashed red line corresponds to the case with $\delta_{0}=0$, $A_{0}=3$,
and $k_{1}=1$, reflecting the center of mass of the $\psi_{1}$ profile
shown in Fig. \ref{Fig.7-up.const-down.periodic}(f). The divergence
of the center of mass in this case from the expected result when considering
only the constant interaction is noteworthy.

However, for values of $k_{1}$ greater than $k_{1}\approx2$, the
system returns to a bound state, as evidenced by the dashed blue line.
It is interesting to note that for $k_{1}=5$ (dash-dotted green line)
and $k_{1}=10$ (yellow circles), with $\delta_{0}=0$ and $A_{0}=3$,
the center of mass of the system converges to the expected value when
only the constant interaction is considered (solid black line). This
trend reveals the system's ability to realign into a bound state when
the wave vector $k_{1}$ is significantly increased, thereby restoring
expected characteristics in the system's dynamics.

In Fig. \ref{Fig.8-a.CM-b.VAR-k1-maior}(b), a similar analysis is
performed for the variance of the solution of the field $\psi_{1}$.
It is notable how the solution's variance decreases as the value of
$k_{1}$ increases, demonstrating a convergence trend towards the
solid black curve, which represents the outcome when only the constant
component of $\delta(z)$ is considered. This behavior illustrates
the system's ability to reduce the solution's dispersion as the wave
vector $k_{1}$ is incremented, achieving a behavior more aligned
with what is expected in the absence of the quasi-periodic modulation.

\begin{figure}[tb]
\centering \includegraphics[width=1\columnwidth]{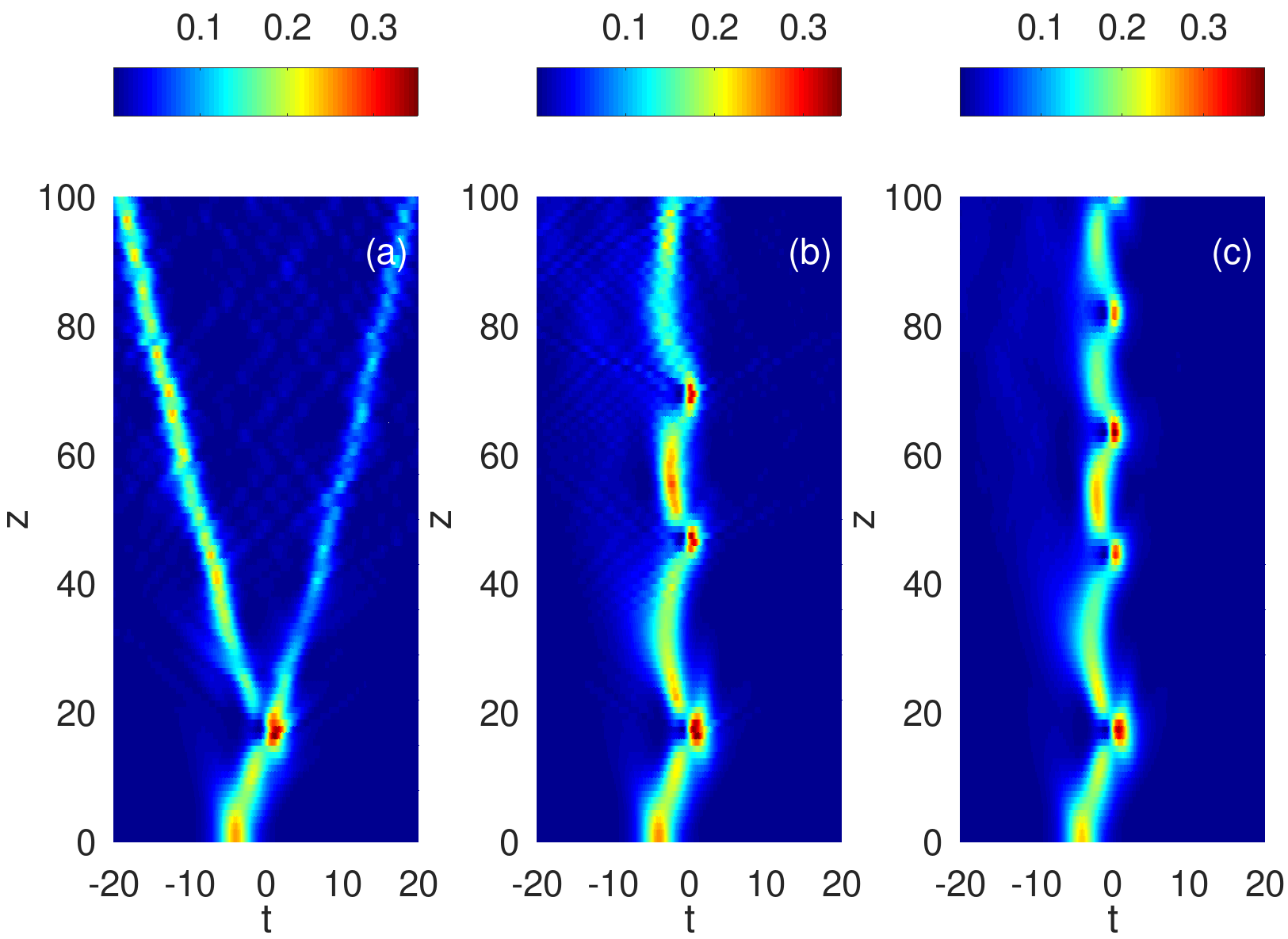}
\caption{Profile of $\psi_{1}$ for a) $k_{1}=1$, b) $k_{1}=2$, and c) $k_{1}=5$
with interaction exclusively due to the quasi-periodic modulation,
$A_{0}=3$. It can be observed that for higher values of $k_{1}$,
the system resembles the behavior of the case with constant interaction,
as represented in Fig. \ref{Fig.7-up.const-down.periodic}(c).}
\label{Fig.9-evol-k1-bigg} 
\end{figure}

To delve deeper into the analysis of the impact of the intensity of
$k_{1}$ on the interaction between the fields, we generated the profile
of $\psi_{1}$ for the system with exclusive quasi-periodic interaction,
as shown in Fig. \ref{Fig.9-evol-k1-bigg}. The system was configured
with $A_{0}=3$ and $\delta_{1}=0$. In Fig. \ref{Fig.9-evol-k1-bigg}(a),
the previously discussed scenario for $k_{1}=1$ is presented. Increasing
this value to $k_{1}=2$ in Fig. \ref{Fig.9-evol-k1-bigg}(b), reveals
a significant change in the solution's dynamics, indicating a return
of the system to a bound state. For $k_{1}=5$ in Fig. \ref{Fig.9-evol-k1-bigg}(c),
a behavior quite similar to the case observed in Fig. \ref{Fig.7-up.const-down.periodic}(c)
is noted, where the system was configured such that the interaction
between the fields was constant.

\section{Conclusion}

\label{sec:4}

The present study investigated the dynamics of solitons in optical
fibers under the influence of quasi-periodic modulation. We explored
scenarios where the interaction between solitons is governed by both
constant and quasi-periodic components of the cross-phase modulation
coefficient $\delta(z)$. Through numerical simulations, we observed
distinct behaviors depending on the parameters $A_{0}$ and $\delta_{0}$,
alongside the wave vectors $k_{1}$ and $k_{2}$.

For $A_{0}>0$, corresponding to attractive interactions, the solitons
tend to approach each other, resulting in a coherent state over propagation
distances. Conversely, for $A_{0}<0$, the solitons exhibit repulsive
behavior, leading to continuous separation along the fiber. The evolution
of soliton profiles, center of mass and variance confirmed these tendencies,
highlighting the significant influence of the quasi-periodic modulation
on soliton dynamics.

Furthermore, varying the wave vector $k_{1}$ revealed interesting
insights. When $k_{1}$ is increased beyond a threshold value, the
system transitions from a dispersed state back to a bound state, resembling
the behavior observed under constant interaction conditions. This
reintegration suggests a critical role of $k_{1}$ in regulating the
soliton interactions and stability in the presence of quasi-periodic
perturbations.

In conclusion, {this work provides a comprehensive
analysis of the influence of quasi-periodic modulation on soliton
dynamics in coupled waveguides, elucidating the conditions under which
bound and unbound states emerge, thereby advancing our understanding
of light pulse control in optical communication systems.}

Future investigations may explore additional complexities and applications
of such systems in nonlinear optics and photonics. They could also
focus on extending the analysis to include higher-order effects of
modulation parameters, such as nonlinear dispersion or higher-order
nonlinearity, to describe more complex soliton dynamics in optical
fibers. Additionally, exploring the impact of different fiber geometries
or external perturbations could provide further insights into optimizing
soliton-based communication systems.

\section*{Fundings}

The authors acknowledge financial support of the Brazilian agencies
CNPq (\#303469/2019-6, \#306105/2022-5, \#312173/2022-9, \#407469/2021-4,
\#402830/2023-7), CAPES and Paraíba State Research Foundation (Pronex
\#0015/2019). This work was also performed as part of the Brazilian
National Institute of Science and Technology (INCT) for Quantum Information
(\#465469/2014-0).

\bibliographystyle{apsrev4-2}
\bibliography{bib}
 
\end{document}